\documentclass[preprint,showpacs,preprintnumbers,amsmath,amssymb,superscriptaddress]{revtex4}

\usepackage{graphicx}
\usepackage{dcolumn}
\usepackage{bm}
\usepackage{color}

\begin{document}

\preprint{APS}

\title{Effect of Pt substitution on the electronic structure of AuTe$_2$}

\author{D.~Ootsuki}
\affiliation{Department of Physics {\rm {\&}} Department of Complexity Science and Engineering, 
University of Tokyo, 5-1-5 Kashiwanoha, Chiba 277-8561, Japan}
\author{K.~Takubo}
\affiliation{Department of Physics {\rm {\&}} Astronomy, University of British Columbia, 
Vancouver, British Columbia V6T\,1Z1, Canada}
\affiliation{Quantum Matter Institute, University of British Columbia, Vancouver, British Columbia V6T\,1Z4, Canada}
\affiliation{Max Planck Institute for Solid State Research, Heisenbergstrasse 1, D-70569 Stuttgart, Germany}
\author{K.~Kudo}
\affiliation{Department of Physics, Okayama University, Kita-ku, Okayama 700-8530, Japan}
\author{H.~Ishii}
\affiliation{Department of Physics, Okayama University, Kita-ku, Okayama 700-8530, Japan}
\author{M.~Nohara}
\affiliation{Department of Physics, Okayama University, Kita-ku, Okayama 700-8530, Japan}
\author{N.~L.~Saini}
\affiliation{Department of Physics, University of Rome "La Sapienza", 00185 Rome, Italy}
\author{R.~Sutarto}
\affiliation{Canadian Light Source, Saskatoon, Saskatchewan  S7N 2V3, Canada}
\author{F.~He}
\affiliation{Canadian Light Source, Saskatoon, Saskatchewan  S7N 2V3, Canada}
\author{T.~Z.~Regier}
\affiliation{Canadian Light Source, Saskatoon, Saskatchewan  S7N 2V3, Canada}
\author{M. Zonno}
\affiliation{Department of Physics {\rm {\&}} Astronomy, University of British Columbia, 
Vancouver, British Columbia V6T\,1Z1, Canada}
\author{M. Schneider}
\affiliation{Department of Physics {\rm {\&}} Astronomy, University of British Columbia, 
Vancouver, British Columbia V6T\,1Z1, Canada}
\author{G.~Levy}
\affiliation{Department of Physics {\rm {\&}} Astronomy, University of British Columbia, 
Vancouver, British Columbia V6T\,1Z1, Canada}
\affiliation{Quantum Matter Institute, University of British Columbia, Vancouver, British Columbia V6T\,1Z4, Canada}
\author{G. A.~Sawatzky}
\affiliation{Department of Physics {\rm {\&}} Astronomy, University of British Columbia, 
Vancouver, British Columbia V6T\,1Z1, Canada}
\affiliation{Quantum Matter Institute, University of British Columbia, Vancouver, British Columbia V6T\,1Z4, Canada}
\author{A.~Damascelli}
\affiliation{Department of Physics {\rm {\&}} Astronomy, University of British Columbia, 
Vancouver, British Columbia V6T\,1Z1, Canada}
\affiliation{Quantum Matter Institute, University of British Columbia, Vancouver, British Columbia V6T\,1Z4, Canada}
\author{T.~Mizokawa}
\affiliation{Department of Physics {\rm {\&}} Department of Complexity Science and Engineering, 
University of Tokyo, 5-1-5 Kashiwanoha, Chiba 277-8561, Japan}
\affiliation{Department of Physics, University of Rome "La Sapienza", 00185 Rome, Italy}

\date{\today}

\begin{abstract}
We report a photoemission and x-ray absorption study on Au$_{1-x}$Pt$_x$Te$_2$ ($x = 0$ and 0.35) 
triangular lattice in which superconductivity is induced by Pt substitution for Au. 
Au $4f$ and Te $3d$ core-level spectra of AuTe$_2$ suggests a valence state of
Au$^{2+}$(Te$_2$)$^{2-}$, which is consistent with its distorted crystal structure 
with Te-Te dimers and compressed AuTe$_6$ otahedra. On the other hand, valence-band photoemission spectra 
and pre-edge peaks of Te $3d$ absorption edge indicate that Au $5d$ bands are almost fully occupied 
and that Te $5p$ holes govern the transport properties and the lattice distortion. 
The two apparently conflicting pictures can be reconciled by strong Au $5d$/Au $6s$-Te $5p$ hybridization.
Absence of a core-level energy shift with Pt substitution is inconsistent with 
the simple rigid band picture for hole doping. 
The Au $4f$ core-level spectrum gets slightly narrow with Pt substitution, 
indicating that the small Au $5d$ charge modulation in distorted AuTe$_2$ is partially suppressed.
\end{abstract}

\maketitle

\section{Introduction}

Layered transition-metal dichalcogenides with triangular motif have been attracting
renewed interest due to the discovery of superconductivity in chemically substituted
IrTe$_2$ \cite{Pyon2012,Yang2012,Kudo2013Ir,Kamitani2013} and AuTe$_2$ \cite{Kudo2013} 
with maximum $T_c$ of 3.1 K and 4.0 K, respectively. 
In particular, the electronic structure of IrTe$_2$ and its derivatives
has been intensively studied using various spectroscopic methods 
under anticipation that the strong spin-orbit interaction in the Ir $5d$ and Te $5p$ 
orbitals may provide a novel spin-momentum entangled quantum state. 
\cite{Ootsuki2012, Fang2012, Ootsuki2013, Oh2013}
Also details of the structural transition in IrTe$_2$ \cite{Jobic1991,Matsumoto1999} 
have been revealed by recent studies using advanced x-ray diffraction and scattering
techniques. \cite{Kiswandhi2013,Joseph2013,Toriyama2014,Pascut2014,Takubo2014}
On the other hand, so far, electronic structure studies on AuTe$_2$ and 
its derivatives are limited although the Au $5d$ and/or Te $5p$ electrons 
with strong spin-orbit interaction can provide an interesting electronic state.

AuTe$_2$ is known as a natural mineral Calaverite with monoclinically 
distorted CdI$_2$-type layered structure (space group C2/m). \cite{Tunell1952}
Each Au-Te layer contains edge-shared AuTe$_6$ octahedra which are strongly distorted 
with two short (2.67 $\AA$) and four long (2.98 $\AA$) Au-Te bonds 
due to Te-Te dimer formation in the average structure. \cite{Tunell1952,Janner1989}
A detailed analysis of the crystal structure has revealed the incommensurate 
structural modulation which may indicate charge ordering of Au $5d$ and/or Te $5p$
valence electrons. \cite{Schutte1988}
Although Au$^+$/Au$^{3+}$ charge disproportionation has been suggested to explain 
the structural distortion, \cite{Schutte1988} the expected Au valence modulation 
has not been detected by x-ray photoemission spectroscopy. \cite{Triest1990}
In addition, {\it ab-initio} calculations have indicated that the Au $5d$ subshell 
is almost fully occupied by electrons and the Te-Te dimer formation due to 
the partially occupied Te $5p$ subshell should be responsible for the structural distortion. 
\cite{Krutzen1990}
Very recently, Kudo {\it et al.} have found that Pt substitution for Au suppresses the lattice 
distortion of AuTe$_2$ and that Au$_{1-x}$Pt$_x$Te$_2$ with undistorted CdI$_2$-type (P$\overline{3}$m1)
structure exhibits superconductivity with maximum $T_c$ of 4.0 K. \cite{Kudo2013} 
The electronic phase diagram for Au$_{1-x}$Pt$_x$Te$_2$ is similar to Ir$_{1-x}$Pt$_x$Te$_2$, 
indicating intimate relationship between the lattice distortion in AuTe$_2$ and 
the superconductivity in Au$_{1-x}$Pt$_x$Te$_2$. 
In the present work, we have studied the fundamental electronic structure of 
Au$_{1-x}$Pt$_x$Te$_2$ ($x$ =0 and 0.35) by means of ultra-violet photoemission spectroscopy (UPS),
x-ray photoemission spectroscopy (XPS), and x-ray absorption spectroscopy (XAS).
The valence-band UPS and XPS results show that the Au $5d$ and Te $5p$ orbitals
are strongly hybridized near the Fermi level.  
The core-level XPS results indicate small charge distribution of the Au $5d$ electrons
which is suppressed by the Pt substitution. Active role of the Te $5p$ holes is indicated by
the Te $3d$ XAS measurement. 

\section{Experiments}
 
The polycrystalline samples of Au$_{1-x}$Pt$_{x}$Te$_2$ (x=0.35, $T_c$=4.0 K) 
and single crystals of AuTe$_2$ were prepared as reported in the literature. \cite{Kudo2013}
Single crystals of AuTe$_2$ were cleaved for UPS and XAS at 300 K. 
UPS measurements were performed at UBC using SPECS Phoibos 150 analyzer with the He I line (21.2 eV) 
from a monochromatized UVS300 lamp. The total energy resolution was set to 25 meV. 
The base pressure was in the 10$^{-11}$ mbar range. 
XAS measurements were performed at beamlines 11ID-1 and 10ID-2
\cite{CLS}, Canadian Light Source. The total-energy resolution was 100 meV. 
The base pressure of the XAS chamber was in the 10$^{-9}$ mbar range. 
The spectra were measured in the total-electron-yield (TEY) mode.
XPS measurements were carried out using JEOL JPS9200 analyzer. 
Mg K$\alpha$ (1253.6 eV) was used as x-ray source. 
The total energy resolution was set to $\sim$ 1.0 eV, and 
the binding energy was calibrated using the Au $4f$ core level 
of the gold reference sample at 84.0 eV. 
The polycrystalline sample of Au$_{1-x}$Pt$_{x}$Te$_2$ (x=0.35) and single crystal 
of AuTe$_2$ were fractured $in$ $situ$ at 300 K for the XPS measurements.

\section{Results and discussion}

\begin{figure}
\includegraphics[width=8cm]{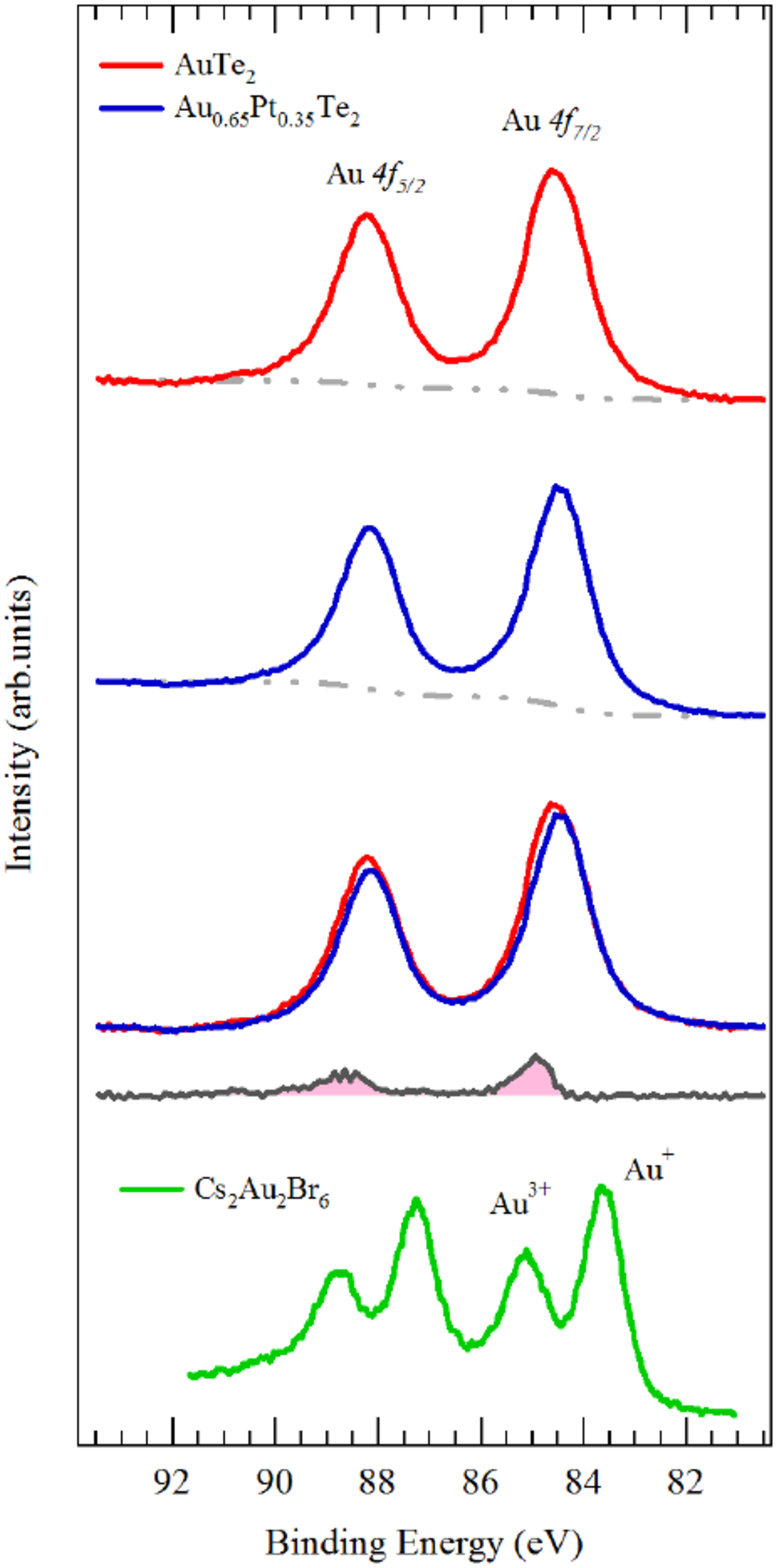}
\caption{(color online)
Au $4f$ core-level XPS spectra of AuTe$_2$ and Au$_{0.65}$Pt$_{0.35}$Te$_2$ 
compared with Cs$_2$Au$_2$Br$_6$. \cite{Son2005} The dash-dot curves indicates 
backgrounds due to secondary electrons. The background subtracted spectra of
Au$_{0.65}$Pt$_{0.35}$Te$_2$ is overlaid with that of AuTe$_2$, and 
the difference spectrum between the background-subtracted spectra
is indicated by the solid curve with shaded peak area.
}
\end{figure}

Figure 1 shows the Au $4f$ core-level spectra of Au$_{1-x}$Pt$_x$Te$_2$ (x=0 and 0.35) 
taken at 300 K, which are compared with Cs$_2$Au$_2$Br$_6$ with Au$^+$ and Au$^{3+}$ sites.
The broad Au $4f_{7/2}$ peak of AuTe$_2$ would be consistent with the Au valence modulation
due to the lattice distortion. However, the Au $4f_{7/2}$ peak width of Au$_{0.65}$Pt$_{0.35}$Te$_2$
without the distortion is also comparable to that of AuTe$_2$.
While the formal valence of Au is +4 in Au$_{1-x}$Pt$_x$Te$_2$, the Au $4f_{7/2}$ peaks 
are slightly higher in binding energy than that of pure Au (84.0 eV) and located 
between the Au$^{+}$ and Au$^{3+}$ peaks of Cs$_2$Au$_2$Br$_6$, suggesting that 
the actual average Au valence in Au$_{1-x}$Pt$_x$Te$_2$ is close to 2+. 
Although the Au$^{2+}$ ion is expected to take the $5d^9$ configuration,
the band-structure calculations on the average structure indicate that 
the Au $5d$ bands are almost fully occupied. \cite{Krutzen1990,Kitagawa2013}

\begin{figure}
\includegraphics[width=8cm]{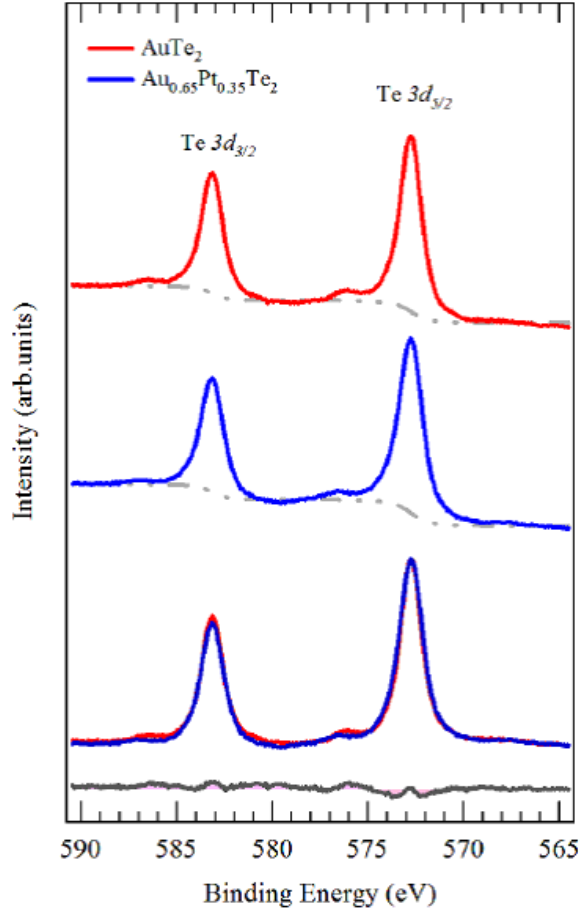}
\caption{(color online)
Te $3d$ core-level XPS spectra of AuTe$_2$ and Au$_{0.65}$Pt$_{0.35}$Te$_2$. 
The dash-dot curves indicates background due to secondary electrons. 
The background subtracted spectra of
Au$_{0.65}$Pt$_{0.35}$Te$_2$ is overlaid with that of AuTe$_2$, and 
the difference spectrum between the background-subtracted spectra
is indicated by the solid curve with shaded peak area.
}
\end{figure}

The Te $3d$ core-level spectra of Au$_{1-x}$Pt$_x$Te$_2$ (x=0 and 0.35) are displayed in Fig. 2. 
The binding energy of the Te $3d_{5/2}$ core level is close to that of pure Te (573.0 eV), \cite{Nyholm1980}
suggesting that the Te $5p$ orbitals are not fully occupied and 
contribute to the electronic states at the Fermi level. 
The shoulder structures located at $\sim$ 576 eV and $\sim$ 587 eV for Te $3d_{5/2}$ 
and Te $3d_{3/2}$ are derived from Te oxide contaminations
which were also observed in the IrTe$_2$ single crystals and 
the Ir$_{1-x}$Pt$_x$Te$_2$ polycrystalline samples. \cite{Ootsuki2012}
The shoulder structures in the AuTe$_2$ single crystal and the polycrystalline
Au$_{1-x}$Pt$_x$Te$_2$ are much smaller than that in the polycrystalline
Ir$_{1-x}$Pt$_x$Te$_2$ and are as small as that in the high quality IrTe$_2$ 
single crystal, indicating that the surface quality of AuTe$_2$ 
and Au$_{1-x}$Pt$_x$Te$_2$ is reasonably good.

In order to clarify the effect of Pt substitution, we have subtracted the core-level spectrum of 
Au$_{0.65}$Pt$_{0.35}$Te$_2$ from that of AuTe$_2$ as displayed in Figs. 1 and 2. 
The Au $4f$ and Te $3d$ core-level peaks do not 
show appreciable energy shift with the Pt substitution. 
The difference spectrum shows that the Au $4f$ core-level spectrum of AuTe$_2$ 
gets slightly narrow with the Pt substitution, while it does not affect 
the Te $3d$ core level appreciably. 
This indicates small Au $5d$ charge modulation in distorted AuTe$_2$ 
and partial suppression of the charge modulation by the Pt substitution. 
Here, one cannot fully exclude the possibility that the Au valence at 
the surface is different from the bulk and that the surface component 
is enhanced in AuTe$_2$. However, the surface condition of the AuTe$_2$ 
single crystal is expected to be better than Au$_{0.65}$Pt$_{0.35}$Te$_2$, 
and the surface component in AuTe$_2$ should be smaller than 
the polycrystalline case if it exists. On the other hand, the Au 4$f$ peak 
is broader in the AuTe$_2$ single crystal than the polycrystalline case.
Therefore, it is natural to assign the extra broadening in AuTe$_2$ 
to the extra charge modulation instead of the surface effect.

In Fig. 3, valence-band XPS and UPS spectra of Au$_{1-x}$Pt$_x$Te$_2$ (x=0 and 0.35) taken at 300 K are displayed.  
The valence-band UPS and XPS spectra of Au$_{1-x}$Pt$_x$Te$_2$ show several structures. 
The broad structures ranging from 0 to 4 eV below the Fermi level can be assigned to the Te $5p$ 
orbitals (mixed with the Au 5$d$/6$s$ orbitals) on the basis of the band-structure calculations on AuTe$_2$. \cite{Krutzen1990, Kitagawa2013} 
The structures from 4.0 to 6.5 eV can be assigned to the Au $5d$ bands since it gains spectral weight 
in going from UPS to XPS, as expected from the photon energy dependence of 
the photoionization cross-section of Au 5$d$ relative to Te 5$p$.
Indeed, the valence-band spectra are consistent with the calculated density of states \cite{Kitagawa2013} 
in which the Au $5d$ bands are located in the region from 4.0 eV to 6.0 eV below the Fermi level.
The valence-band spectra of Au$_{0.65}$Pt$_{0.35}$Te$_2$ is shifted 
toward lower binding energy, indicating that the Pt substitution for Au 
may correspond to hole doping to the Te $5p$ bands mixed 
with the Au $5d$/$6s$ orbitals. 
Another possibility is that mixing of the Au $5d$ bands with the Pt $5d$ bands leads to the energy shift 
of the Au/Pt $5d$ bands since the Pt $5d$ bands are expected to have lower binding energy than the Au $5d$ bands. 
The absence of the core-level energy shift is inconsistent with the former scenario 
(hole doping in a rigid band manner) and supports the latter scenario. 
Namely, the Pt substitution changes the shape of the valence band constructed from the Au/Pt $5d$/6$s$ and Te 5$p$ orbitals,
and it cannot be viewed as a simple hole doping to AuTe$_2$ in a rigid band manner.

The average Au valence close to +2 and the unoccupied Te $5p$ orbitals indicate that 
the charge-transfer energy from the Te $5p$ orbitals to the Au $5d$ orbitals is negative
to stabilize the valence state of Au$^{2+}$(Te$_2$)$^{2-}$.
Namely, the local electronic configuration of the AuTe$_6$ octahedron is close to $d^9\underbar{L}^2$ 
($\underbar{L}$ represents a ligand hole in the Te $5p$ orbitals) instead of $d^7$. Therefore, each Te site 
accommodates approximately one hole, and the Te $5p$ holes govern the transport properties 
and the lattice distortions in AuTe$_2$. This picture is consistent with the Te-Te dimers
in AuTe$_2$ since the antibonding molecular orbital of the Te-Te dimer can be occupied 
by the two Te $5p$ holes from the two Te sites.
On the other hand, the band-structure calculations on AuTe$_2$ with the average structure 
\cite{Krutzen1990,Kitagawa2013}
as well as the valence-band spectra indicate that the Au $5d$ bands are almost fully occupied. 
In order to resolve this apparent paradox, strong hybridization between the Au $5d$/6$s$ 
and Te $5p$ orbitals should be taken into account.
Starting from the Au$^{2+}$(Te$_2$)$^{2-}$ valence state, the hybridization between the Au 5$d$/6$s$ 
orbitals and the Te-Te bonding and antibonding molecular orbitals can induce additional charge transfer. 
Since the Au $5d$ level in AuTe$_2$ is much lower than the Ir $5d$ level in IrTe$_2$, charge donation 
from the Te-Te bonding orbital to the Au $5d$ orbitals can be dominant in AuTe$_2$ 
whereas back donation from the Ir 5$d$ orbitals to the Te-Te antibonding orbital would be 
substantial in IrTe$_2$. 
In addition, the Au 6$s$ component can be mixed into the Au 5$d$ bands through the strong
Au 5$d$-Te 5$p$ and the Te 5$p$-Au 6$s$ hybridizations. Therefore, although the "Au 5$d$ bands" 
constructed from the atomic Au $5d$, Au 6$s$, and Te 5$p$ orbitals are fully occupied as predicted by 
the band-structure calculations and observed by the valence-band photoemission experiments, 
actual number of atomic Au 5$d$ electrons in AuTe$_2$ can remain close to nine which is
consistent with the Au$^{2+}$(Te$_2$)$^{2-}$ valence state.
The $d^9$ configuration of Au$^{2+}$ is consistent with the Jahn-Teller like distortion of the AuTe$_6$ octahedra 
with two short and four long Au-Te bonds.

\begin{figure}
\includegraphics[width=8cm]{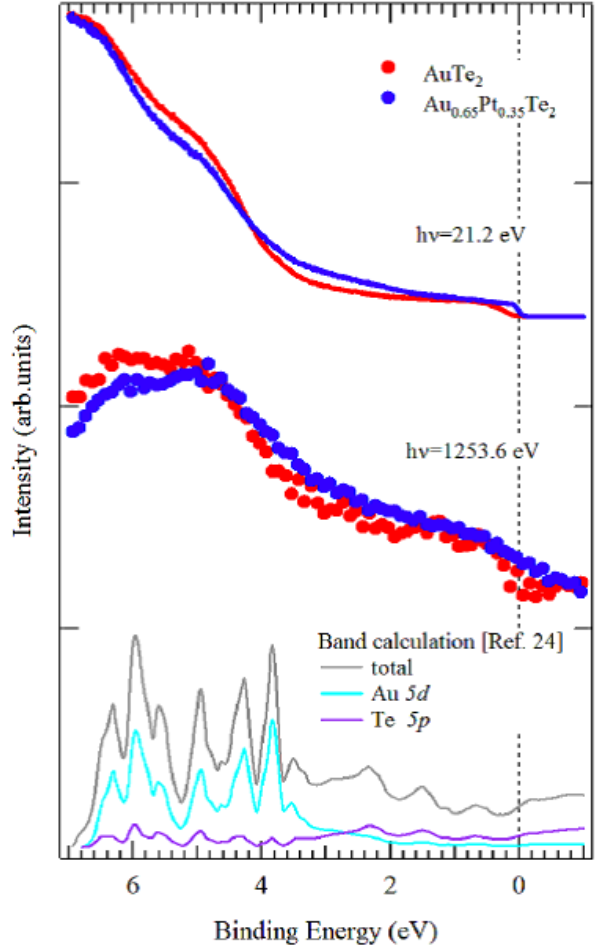}
\caption{(color online)
Valence-band UPS and XPS spectra of AuTe$_2$ and Au$_{0.65}$Pt$_{0.35}$Te$_2$ 
compared with the total and partial density of states of AuTe$_2$. \cite{Kitagawa2013}
}
\end{figure}

\begin{figure}
\includegraphics[width=10cm]{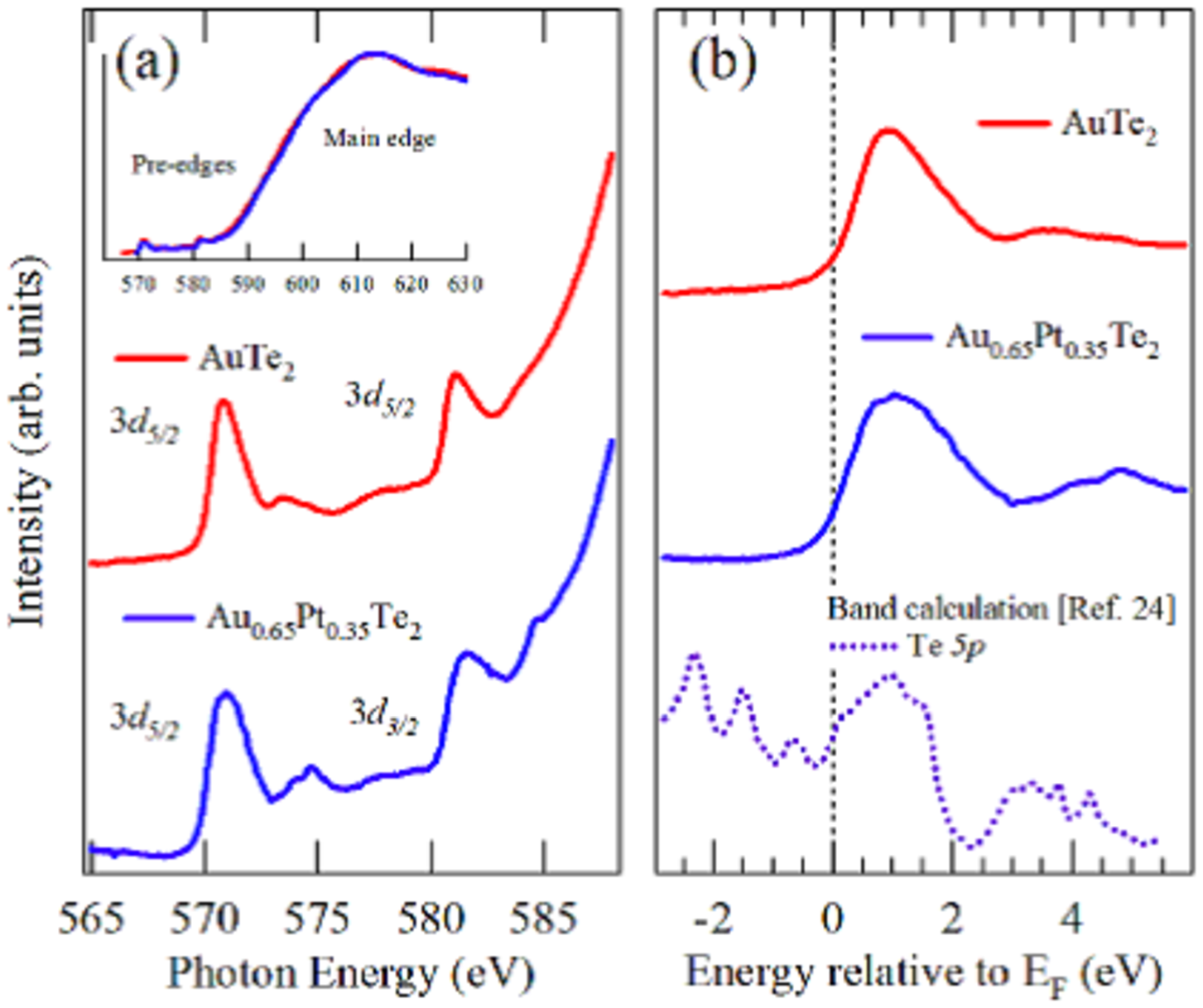}
\caption{(color online)
(a) Te $3d$ XAS spectrum of AuTe$_2$ and Au$_{0.65}$Pt$_{0.35}$Te$_2$. 
The inset shows the wide-range XAS spectrum including the main edge. 
(b) The Te 3$d_{5/2}$ XAS spectrum is compared with the calculated density of states. \cite{Kitagawa2013} 
Here, it is assumed that the Fermi level is roughly located around the absorption edge.}
\end{figure}

Figure 4 shows the Te $3d$ XAS spectrum of Au$_{1-x}$Pt$_x$Te$_2$ (x=0 and 0.35). 
The pre-edge and main edge structures are clearly observed. 
The main-edge structure corresponds to transition from the Te $3d$ core level
to the unoccupied Te $4f$/Au $6s$,$6p$ states. 
On the other hand, the pre-edge structure can be assigned to the transition from
the Te $3d$ core level to the Te $5p$ orbitals, 
indicating that the Te $5p$ bands cross the Fermi level and that the Te $5p$ holes play essential roles in the transport properties. 
This Te $5p$-hole picture is consistent with the XPS and XAS results. 
In AuTe$_2$, the Te $5p$ orbitals are partially unoccupied, and the bond formation by the Te $5p$ holes creates the Te-Te dimers. 
The Te-Te dimer formation leads to the long and short Te-Te bonds, \cite{Tunell1952,Janner1989} which can induce 
charge modulation of Au through the strong  hybridization between the Au $5d$ and Te $5p$ orbitals. 
On the other hand, since all the Te sites belong to one of the Te-Te dimers, each Te site accommodates almost same amount of Te $5p$ hole.
When Pt is substituted for Au in Au$^{2+}$(Te$_2$)$^{2-}$, Pt ions tend to be 3+ or 4+ and supply electrons to Te-Te antibonding orbitals. 
Consequently, the local Te-Te dimers are partly broken around the Pt sites, 
and the superstructure due to the short Te-Te bond (intradimer) and 
long Te-Te bond (interdimer) is strongly disturbed.
In this scenario, disordered local Te-Te dimers can remain in Au$_{1-x}$Pt$_x$Te$_2$. 
The remaining Au charge fluctuation and the disordered Te-Te dimers may provide 
anomalous lattice behaviors to Au$_{1-x}$Pt$_x$Te$_2$ and may contribute to 
the emergence of superconductivity. \cite{Kudo2013}

\section{Conclusion}

We have performed photoemission and x-ray absorption measurements on Au$_{1-x}$Pt$_x$Te$_2$ 
($x$ =0 and 0.35) in which the Pt substitution for Au suppresses the lattice distortion in AuTe$_2$ and induces the superconductivity.
The broad Au $4f$ core-level peak is consistent with the Au valence modulation in distorted AuTe$_2$.
The Au $4f$ core-level peak gets slightly narrow with the Pt substitution, indicating that small Au 5$d$ 
charge modulation in distorted AuTe$_2$ is at least partially suppressed by the Pt substitution.
The Au $4f$ and Te 3$d$ core-level binding energies suggest that average valence state is close to Au$^{2+}$(Te$_2$)$^{2-}$ 
consistent with the Jahn-Teller like distortion of the AuTe$_6$ octahedra.
On the other hand, the valence-band spectra and the band-structure calculations show that the Au 5$d$ bands
are almost fully occupied. The two apparently conflicting results can be reconciled by taking account of
the strong Au 5$d$/Au 6$s$-Te 5$p$ hybridization.
The absence of core-level energy shift with the Pt substitution shows that the simple rigid band picture is not applicable
to Au$_{1-x}$Pt$_x$Te$_2$. Although the periodic arrangement of the Te-Te dimers is disturbed by the Pt substitution,
the Te-Te dimers and Au valence modulation may partly remain in superconducting Au$_{1-x}$Pt$_x$Te$_2$.
The relationship between the possible Au charge fluctuation and the superconductivity should be studied experimentally 
and theoretically in future. Another interesting question is that the Te-Te dimers still remain 
in Au$_{1-x}$Pt$_x$Te$_2$ or not. If the Pt substitution causes disordering of the dimers instead of breaking,
Au$_{1-x}$Pt$_x$Te$_2$ should have highly inhomogeneous electronic states similar to the Fe-based superconductors. 

\section*{Acknowledgements}

The authors would like to thank Professor Y. Ohta and Dr. T. Toriyama for informative discussion.
This work was partially supported by Grants-in-Aid from the Japan Society of 
the Promotion of Science (JSPS) (Grants No: 22540363, 25400372, 25400356, and 26287082) 
and the Funding Program for World-Leading Innovative R\&D on Science and Technology 
(FIRST Program) from JSPS.
D.O. acknowledges supports from the JSPS Research Fellowship for Young Scientists.
The work at UBC was supported by the Max Planck - UBC Centre for Quantum Materials, 
the Killam, Alfred P. Sloan, Alexander von Humboldt, and NSERC's Steacie Memorial Fellowships (A.D.), 
the Canada Research Chairs Program (A.D., G.A.S.), NSERC, CFI, and CIFAR Quantum Materials. 
The XAS experiments were carried out at beamline 11ID-1 and 10ID-2, 
Canadian Light Source (Proposals ID: 16-4388 and 18-5295).
Part of the research described in this paper was performed at the Canadian Light Source, 
which is funded by the CFI, NSERC, NRC, CIHR, the Government of Saskatchewan, 
WD Canada, and the University of Saskatchewan.


\begin{thebibliography}{99}

\bibitem{Pyon2012}
S. Pyon, K. Kudo, and M. Nohara, J. Phys. Soc. Jpn. {\bf 81}, 053701 (2012).

\bibitem{Yang2012}
J. J. Yang, Y. J. Choi, Y. S. Oh, A. Hogan, Y. Horibe, K. Kim, B. I. Min, and S-W. Cheong,
Phys. Rev. Lett. {\bf 108}, 116402 (2012).

\bibitem{Kudo2013Ir}
K. Kudo, M. Kobayashi, S. Pyon, and M. Nohara, J. Phys. Soc. Jpn. {\bf 82}, 085001 (2013).

\bibitem{Kamitani2013}
M. Kamitani, M. S. Bahramy, R. Arita, S. Seki, T. Arima, Y. Tokura, and S. Ishiwata, Phys. Rev. B {\bf 87}, 180501(R) (2013).

\bibitem{Kudo2013}
K. Kudo, H. Ishii, M. Takasuga, K. Iba, S. Nakano, J. Kim, A. Fujiwara, and M. Nohara,
J. Phys. Soc. Jpn. {\bf 82}, 063704 (2013).

\bibitem{Ootsuki2012}
D. Ootsuki, Y. Wakisaka, S. Pyon, K. Kudo, M. Nohara, M. Arita, H. Anzai, H. Namatame, M. Taniguchi, 
N. L. Saini, and T. Mizokawa, Phys. Rev. B {\bf 86}, 014519 (2012).

\bibitem{Fang2012}
A. F. Fang, G. Xu, T. Dong, P. Zheng, and N. L. Wang, Sci. Rep. {\bf 3}, 1153 (2013). 

\bibitem{Ootsuki2013}
D. Ootsuki, S. Pyon, K. Kudo, M. Nohara, M. Arita, H. Anzai, H. Namatame, M. Taniguchi, 
N. L. Saini, and T. Mizokawa, J. Phys. Soc. Jpn. {\bf 82}, 093704 (2013). 

\bibitem{Oh2013}
Yoon Seok Oh, J. J. Yang, Y. Horibe, and S.-W. Cheong, Phys. Rev. Lett. {\bf 110}, 127209 (2013).

\bibitem{Jobic1991}
S. Jobic, P. Deniard, R. Brec, J. Rouxel, A. Jouanneaux, and A. N. Fitch, 
Z. Anorg. Allg. Chem. {\bf 598}, 199 (1991).

\bibitem{Matsumoto1999}
N. Matsumoto, K. Taniguchi, R. Endoh, H. Takano, and S. Nagata, J. Low Temp. Phys. {\bf 117}, 1129 (1999).
 
\bibitem{Kiswandhi2013}
A. Kiswandhi, J. S. Brooks, H. B. Cao, J. Q. Yan, D. Mandrus, Z. Jiang, and H. D. Zhou, 
Phys. Rev. B {\bf 87}, 121107(R) (2013).

\bibitem{Joseph2013}
B. Joseph, M. Bendele, L. Simonelli, L. Maugeri, S. Pyon, K. Kudo, M. Nohara, T. Mizokawa, and N. L. Saini, 
Phys. Rev. B {\bf 88}, 224109 (2013).

\bibitem{Toriyama2014}
T. Toriyama, M. Kobori, Y. Ohta, T. Konishi, S. Pyon, K. Kudo, M. Nohara, K. Sugimoto, T. Kim, and A. Fujiwara, 
J. Phys. Soc. Jpn. {\bf 83}, 033701 (2014).

\bibitem{Pascut2014}
G. L. Pascut, K. Haule, M. J. Gutmann, S. A. Barnett, A. Bombardi, S. Artyukhin, T. Birol, D. Vanderbilt, 
J. J. Yang, S.-W. Cheong, and V. Kiryukhin, Phys. Rev. Lett. {\bf 112}, 086402 (2014).

\bibitem{Takubo2014}
K. Takubo, R. Comin, D. Ootsuki, T. Mizokawa, H. Wadati, Y. Takahashi, G. Shibata, A. Fujimori, R. Sutarto, F. He,
S. Pyon, K. Kudo, M. Nohara, G. Levy, I. Elfimov, G. A. Sawatzky, A. Damascelli, arXiv:1405.7766.

\bibitem{Tunell1952}
G. Tunell and L. Pauling, Acta Cryst. {\bf 5}, 375 (1952).

\bibitem{Janner1989}
A. Janner and B. Dam, Acta Crystallogr. Sect. A {\bf 45}, 15 (1989).

\bibitem{Schutte1988}
W. J. Schutte and J. L. de Boer,
Acta Crystallogr. Sect. B {\bf 44}, 486 (1988).

\bibitem{Triest1990}  
A. van Triest, W. Folkerts, and C. Haas, J. Phys.: Condens. Matter {\bf 2}, 8733 (1990). 

\bibitem{Krutzen1990}  
B. C. H. Krutzen and J. E. Inglesfield: J. Phys.: Condens. Matter {\bf 2}, 4829 (1990).

\bibitem{CLS}
D. G. Hawthorn, F. He, L. Venema, H. Davis, A. J. Achkar, J. Zhang, R. Sutarto, H. Wadati, A. Radi, 
T. Wilson, G. Wright, K. M. Shen, J. Geck, H. Zhang, V. Novk, and G. A. Sawatzky, Rev. Sci. Instrum. {\bf 82}, 073104 (2011).

\bibitem{Son2005}
J.-Y. Son, T. Mizokawa, J. W. Quilty, K. Takubo, K. Ikeda, and N. Kojima,
Phys. Rev. B {\bf 72}, 235105 (2005).

\bibitem{Kitagawa2013}
S. Kitagawa, H. Kotegawa, H. Tou, H. Ishii, K. Kudo, M. Nohara, H. Harima, 
J. Phys. Soc. Jpn. {\bf 82}, 113704 (2013). 

\bibitem{Nyholm1980}
R. Nyholm and N. M\r{a}rtensson,
J. Phys. C: Solid State Phys. {\bf 13}, L279 (1980).

\end{thebibliography}
\end{document}